\documentclass{aa}  %[referee]--colocar antes do aa
\usepackage{graphicx}
\usepackage{natbib}
\usepackage{subfig}

\bibpunct{(}{)}{;}{a}{}{,}
\def\tam{0.675}

\begin{document}
\title{Spiral-like structure at the centre of nearby clusters of galaxies}
\author{T. F. Lagan\'{a} \inst{1}
\and F. Andrade-Santos\inst{1}
\and G. B. Lima Neto\inst{1}}

\institute{Universidade de S\~{a}o Paulo, Instituto de Astronomia, Geof\'{i}sica e Ci\^{e}ncias Atmosf\'{e}ricas, Departamento de Astronomia, Rua do Mat\~{a}o 1226, Cidade Universit\'{a}ria,
05508-090, S\~{a}o Paulo, SP, Brazil.}

\date{Received  26 August 2009 / Accepted 17 November 2009}

\abstract
{X-ray data analysis have found that fairly complex structures at cluster centres are more  
common than expected. Many of these structures have similar morphologies, which exhibit spiral-like substructure.}
{It is not yet well known how these structures formed or are maintained.
Understanding the origin of these spiral-like features at the centre of some clusters
is the major motivation behind this work.}
{We analyse deep \textit{Chandra} observations of 15 nearby galaxy clusters (0.01 $ < z < $ 0.06), and
use X-ray temperature and substructure maps to detect small features at the cores of the clusters.}
{We detect spiral-like features at the centre of 7 clusters: A85, A426, A496, Hydra A cluster, Centaurus, Ophiuchus,
 and A4059. These patterns are similar to 
those found in numerical hydrodynamic simulations of cluster mergers with non-zero impact parameter. 
In some clusters of our sample, a strong radio source also occupies the inner region of the cluster, which indicates 
a possible connection between the two. 
Our investigation implies that these spiral-like structures may be caused by off-axis minor mergers. 
Since these features occur in regions of high density, they may confine radio emission from 
the central galaxy producing, in some cases, unusual radio morphology.}
{}

\maketitle 

\section{Introduction}
Previous studies have shown that many clusters continue to be in the process of forming, groups or individual galaxies 
being accreated from the outer large scale structure filaments \citep[e.g.,][]{Durret03}. Temperature maps of the X-ray emitting gas have 
shown that even clusters with apparently relaxed structures in X-ray can have perturbed temperature maps, 
which are indicative of recent ongoing mergers \citep[e.g.,][]{Finoguenov05,Durret08}. The detection of these small structures 
is one of the main achievements of the present era of X-ray telescopes such as {\it Chandra} and XMM-{\it {\textit Newton}}.
However, the presence of cooling cores suggest that recent merger(s) have not had enough time to destroy the cooling core. 

More than 70\% of cooling-flow clusters contain a central cD galaxy that is a radio source 
\citep[e.g.,][]{Burns90,Eilek04}.\textit{Chandra} observations of these systems 
found that the core of many clusters exhibit morphological complexities that are 
probably produced by the interaction between the intracluster medium (ICM) and 
the central radio galaxy. 

The most important connection between the X-ray ICM and the non-thermal
relativistic electrons that produce radio emission is dynamical. 
In the case of very energetic jets, the ICM is heated and compressed because of supersonic shocks
\citep{Heinz98,Reynolds01}. However, there are weaker jets that do not provide an efficient shock
heating, but instead a bubble of low-density gas, which can be identified as a depression in the 
X-ray surface brightness, that rises because of buoyancy \citep[e.g.,][]{Churazov00,Bruggen02,McNamara05}. 

On the other hand, the ICM can also affect the radio emission from radio galaxies by placing upper limits on 
the radio emission. The unusual radio morphology of central cluster galaxies that fill X-ray cavities 
of low density has been reported previously \citep{Bohringer93,Taylor94,Fabian00,Fabian02,Taylor02}. The most extreme case of
radio distortion is the amorphous radio galaxy PKS 0745-191 at the core of the Centaurus cluster \citep{Taylor94}. 

Another interesting feature is the spiral-like feature at the core of Perseus \citep{Churazov03,Fabian00,Taylor02}, A2204 \citep{Sanders08}, and A2029 \citep{Clarke04}.
This spiral pattern was also found in hydrodynamical simulations of merging clusters with non-zero 
impact parameters \citep{Gomez02,Ascasibar06}. In these simulations, the infalling sub-cluster causes a 
disturbance in the mass peak as it passes through the cluster centre. Thus, the central cool gas of the main 
cluster acquires angular momentum, 
producing its spiral pattern.

To understand more clearly the ICM-radio interaction and to characterize the spiral feature, 
we present a sample of 15 nearby clusters, among which 7 exhibit
this feature in temperature and substructure maps, and a large cooling-flow and an embedded radio galaxy at the centre. 
The paper is organized as follows: in Sect.~\ref{sample}, we describe the data sample and the {\it Chandra} 
data analysis; in Sect.~\ref{imaging}, we explain the image procedures used to detect the spiral-like
pattern; in Sect.~\ref{res}, we
present our results and notes for individual clusters; 
in Sect.~\ref{disc}, we discuss a possible scenario to explain the spiral-like
feature formation and the ICM-radio connection; in Sect.~\ref{SyntImg}, we present a synthetic image
produced to reproduce the surface brightness of the Perseus cluster, 
and in Sect.~\ref{conc}, we summarize our findings.

All distance-dependent quantities are derived by assuming the Hubble constant
$H_{0}=70$~km~s$^{-1}$~Mpc$^{-1}$, $\Omega_{M}=0.3$ and, $\Omega_{\Lambda}$= 0.7.

\section{Observation and data reduction}
\label{sample}

\subsection{Cluster selection}

The objects in our sample were selected from a set of clusters within the redshift range 
0.01 $ < z < $ 0.06 with {\it Chandra} public data. We imposed that the data for each clusters had a long exposure time, 
that is, of at least 35 ks. Our final sample contained 15 clusters, whose properties are presented in Tab.~\ref{infos}.

\begin{table*}[ht!]
\centering
\caption{Observational information}
\begin{tabular}{ccccccccc}
\hline\hline
Cluster & RA  & Dec & z & Detector & Exposure time & Soft band$^{\ast}$  & Hard band$^{\ast}$ & $n_{\rm H}^{\ddag}$\\
   & (J2000) & (J2000) &  &  & (ks)& (keV) & (keV) & ($10^{20} ~\rm cm^{-2}$)\\
\hline
Abell 85 	&	00:41:37.80     &  -09:20:33.0 	& 0.055 & ACIS-I & 38.91 & 0.3-1.4 & 1.4-6.0  & 2.78\\
Abell 426       &	03:19:48.20	&  +41:30:42.2	& 0.018	& ACIS-S & 98.20 & 0.3-1.4 & 1.4-6.0  &13.6 \\
Abell 496       &	04:33:37.00	& -13:14:17.0	& 0.033	& ACIS-S & 76.08 & 0.3-1.3 & 1.3-6.0  & 3.78 \\
Abell 3376      &	06:02:10.00	& -39:57:21.0	& 0.046	& ACIS-I & 44.85 & 0.3-1.5 & 1.5-6.0  & 4.58\\
Abell  754      &	09:09:09.00	& -09:39:39.0	&0.054	& ACIS-I & 44.77 & 0.3-1.5 & 1.5-6.0  & 4.82\\
Hydra A         &	09:18:05.70	& -12:05:42.5	&0.013	& ACIS-S & 100.13 & 0.3-1.2 & 1.2-6.0 & 4.60\\
Centaurus	&	12:48:48.90	& -41:18:44.4	&0.011	& ACIS-S & 90.19 & 0.3-1.2 & 1.2-6.0  & 8.56\\
Abell 1644      &	12:57:33.00	& -17:20:28.0	&0.047	& ACIS-I & 52.17 & 0.3-1.6 & 1.6-6.0  & 4.11\\
Abell 1991      &	14:54:31.50	& +18:38:32.0	&0.059	& ACIS-S & 38.81 & 0.3-1.1 & 1.1-6.0  & 2.46 \\
Abell 2052      &	15:16:44.50	& +07:01:16.6	&0.035	& ACIS-S & 128.63 & 0.3-1.2 & 1.2 6.0 & 2.71\\
Abell 2107      &	15:39:39.00	& +21:46:58.0	&0.041	& ACIS-I & 36.04 & 0.3-1.5 & 1.5-6.0  & 4.45\\
Ophiuchus	&	17:12:27.80	& -23:22:11.5	&0.028	& ACIS-S & 51.18 & 1.3-4.0 & 4.0-7.0  & 40.0\\
Abell 3667      &	20:13:07.25	& -56:53:24.0	&0.056  & ACIS-I & 105.01 & 0.3-1.6 & 1.6-6.0 & 4.44\\
Abell 4059      &	23:57:00.70	& -34:45:33.0	&0.047	& ACIS-S & 93.34 & 0.3-1.2 & 1.2-6.0  & 1.21\\
Abell 2589      &	23:23:57.40	& +16:46:39.0	&0.041	& ACIS-S & 54.13 & 0.3-1.3 & 1.3-6.0  & 3.15 \\
\hline
\hline	 
\end{tabular}
\\
\begin{minipage}{15.0truecm}
\footnotesize{$^\ast$ Soft and hard bands used to construct the hardness ratio.}\\
\footnotesize{$\ddag$ Hydrogen column densities adopted to construct HR \citep{DL90}.}
\end{minipage}
\label{infos}
\end{table*}

\subsection{\textit{Chandra} data reduction}

To obtain calibrated images without artifacts, that are adequate for achieving
our goals, it is necessary to follow a series of procedures, otherwise we would have contamination,
which can be detected as substructures.

We used the package CIAO 3.4.
A level 2 event file was initially generated from a level 1 event file,
using the standard procedure\footnote{http://cxc.harvard.edu/ciao3.4/threads/createL2/} and
the latest calibration of CALDB 3.3.0. Periods with flares, which are
spurious high counts caused by protons accelerated by the Sun, were excluded using
the \textit{lc$\_$clean} script. At this point, a rebinned image with pixels corresponding 
to 16 raw physical pixels (4x4) was created from the new level 2 event file, 
in the energy band from 0.3 to 7.0 keV.
We then produced exposure maps and used them to obtain flat images from which the source points
were removed by filling circles around each source by randomly sampling the same
distribution as found in a circular region close to the source. 
Finally, we fitted a 2D analytical surface brightness model.

\section{Imaging}
\label{imaging}
Since the bremsstrahlung emissivity depends on both the temperature and surface brightness, 
irregularities can be detected either in terms of temperature or density discontinuities.
Since $\epsilon \propto T^{1/2}~n^{2}$, these substructures are more significant in the
surface brightness maps. Thus, to resolve the degeneracy, we used substructure 
\citep[which depends on the surface brightness distribution,][]{AndradeSantos09} and 
temperature maps to analyse in detail the inner parts of the clusters in our sample.
of significant structure in the inner region of several clusters.
To investigate this further, we compared these images with the temperature and the substructure maps.

\subsection{X-ray substructures}
\label{substruc}
We developed a new method to quantify substructures in clusters of galaxies using 2D surface brightness fits. 
This method was tested for 47 galaxy clusters observed with \textit{Chandra} \citep{AndradeSantos09}.

The method is based on analysing the number and intensity of substructures detected by a
\textit{friends-of-friends} (FOF) algorithm. This analysis is performed on a residual image, which is 
produced by fitting a bidimensional analytical model ($\beta$-model or S\'ersic) with elliptical symmetry (using Sherpa/CIAO, which fits a
model by minimizing the $\chi^2$) to the cluster X-ray image. 
All surface brightness model parameters are free to vary, including the elipticity, 
position angle, and centre. 
In particular, to identify substructures, we used a threshold in the residual image 
to separate the pixels that had counts statistically significantly 
above or below the 2D fitted model surface brightness at the corresponding position. 
The threshold was defined to be 3 times the local X-ray background variance. Once the pixels 
were separated, we identified these regions as substructures, using the FOF algorithm 
to link together the pixels above the threshold. 
This is a robust method if applied to high signal to noise ratio observations,
which is the case, since the clusters in our sample are rich objects at very low redshifts
($z < 0.06$) and  were observed for more than 35ks.

\subsection{Temperature maps}
\label{kTmap}
Ideally, one would prefer to compute a temperature map using spatially resolved 
spectroscopy as we did in \citet{Lagana08}. This, however, is very timing consuming. 
We opt for another approach, namely to compute the hardness ratio maps and then converte 
them to temperature map. In this way, we are able to detect significant small-scale 
features without resorting to a computer intensive method.

The hardness ratio (HR) maps, or color maps, are defined as the ratio of the fluxes 
in two different bands and can be interpreted as a projected temperature map. 
They can be defined as [smoothed hard-energy band image - smoothed soft-energy band image] / 
[smoothed hard-energy band image + smoothed soft-energy band image]. 
These bands are specified for 
each cluster in Table~\ref{infos} and were chosen such that the two bands had 
almost identical net counts.

We used the {\it csmooth} tool in {\it Chandra} Interactive Analysis of Observations (CIAO), 
an adaptive smoothing algorithm, 
to bin the data using bins of large angular size in order to examine possible structures in the image.
In adaptive binning, the size of the smoothing kernel changes over the original image to create a 
constant S/N per pixel in the output image. The hard and soft energy band  images 
were smoothed identically, with the same adaptive kernel, avoiding spurious artifacts 
and yielding meaningful hardness ratios. 

\begin{figure}[ht!]
\centering
\includegraphics[width=0.45\textwidth]{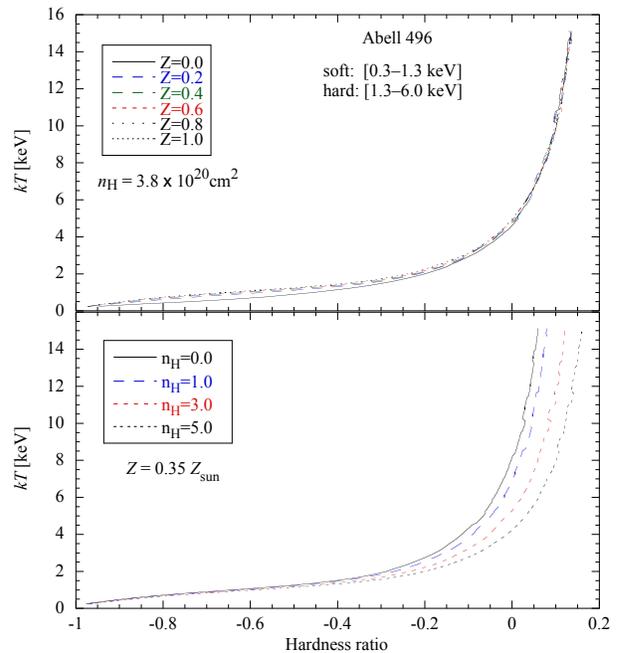}
\caption[]{Hardness ratio conversion into plasma temperature, using Abell 496 as an example. \textsf{Top}: 
HR-to-temperature conversion assuming different metallicity values (in solar units) 
when all other parameters remained fixed. 
\textsf{Bottom}: HR-to-temperature conversion assuming different hydrogen column densities 
(in $10^{20}$~cm$^{-2}$). 
While a metallicity error has an insignificant effect, $n_{\rm H}$ is important for an accurate determination of 
$kT$ (in particular at high temperatures).}
\label{hr2kt}
\end{figure}

To convert the HR maps into temperature maps, we required a simple algorithm to 
compare the relative fluxes in different X-ray bands with a theoretical plasma emission model. 
Thus, based on the assumption of a single temperature, the HR-kT relation was estimated for each cluster, 
and an example is shown in Fig.~\ref{hr2kt} for A496. 
\citet{Watanabe01} also used this procedure to construct the temperature map of 
Ophiuchus using ASCA data.
To do so, we took into account the weighted reponse matrices (RMFs), 
affective area files (ARFs) generated with CIAO 3.4., exposure maps of each observation, 
the hydrogen column densities of each cluster (as specified in Table~\ref{prop}), and adopted a
fixed metallicity value of 0.35 for all clusters. In Fig.~\ref{hr2kt}, we show that the HR conversion into
plasma temperature exhibits little variation assuming different metallicity values.

\section{Results}
\label{res}

In Fig.\ref{HR}, we present the X-ray images, temperature and substructure maps 
of the 15 clusters in our sample. To investigate the properties of these clusters, 
we also present in Tab.\ref{prop} the mean X-ray temperature, the cooling-flow time, 
and the X-ray luminosity 
obtained from the literature. 

\begin{figure*}[ht!]
\centering
\includegraphics[width=\tam\textwidth]{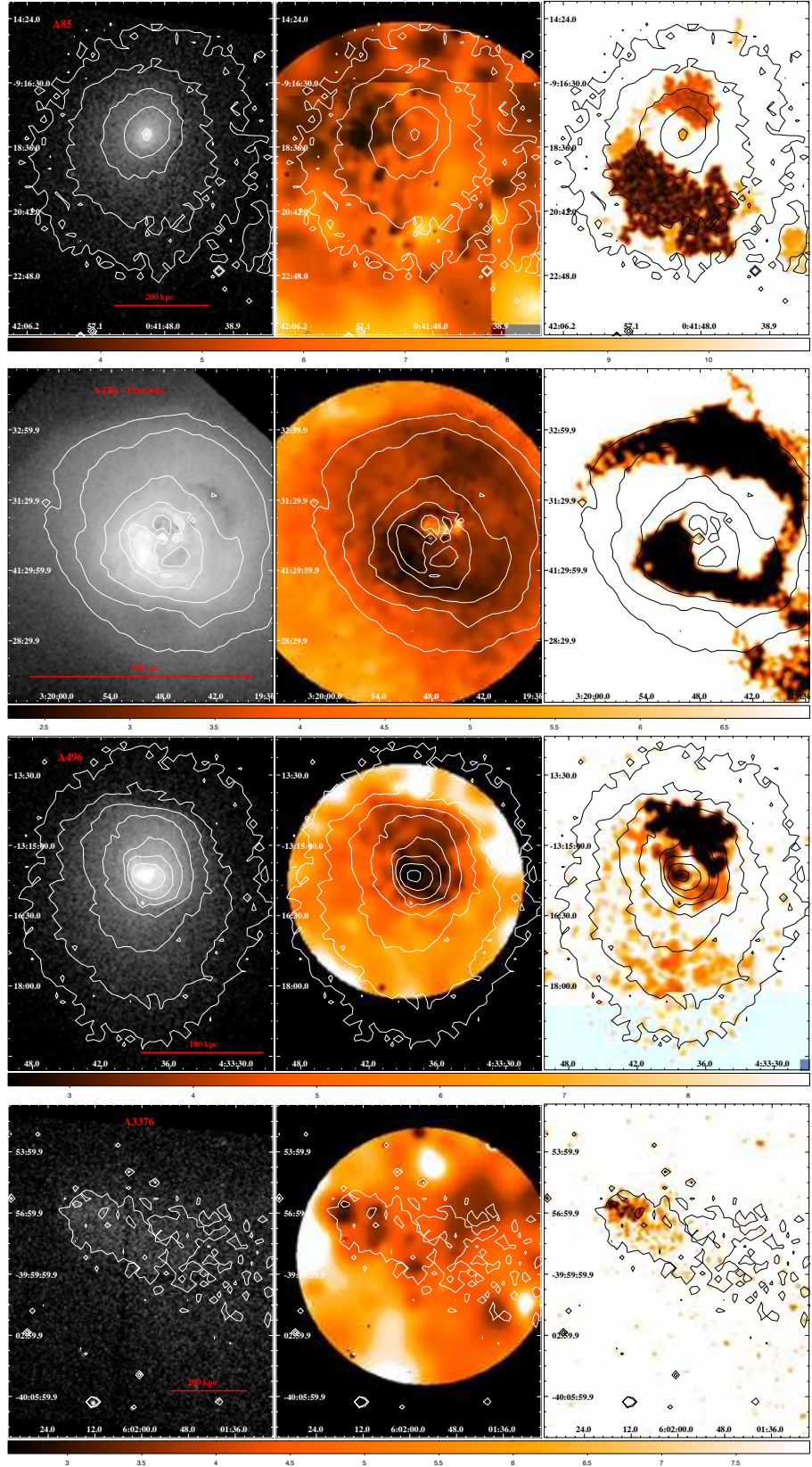}
\caption{Left panels: X-ray image; middle panel: temperature maps; and right panels: substructure maps for
A85, A426 (Perseus), A496, and A3376. The color-bar indicates the temperature in keV.}
\label{HR}
\end{figure*}

\begin{figure*}[ht!]
\centering
\ContinuedFloat
\includegraphics[width=\tam\textwidth]{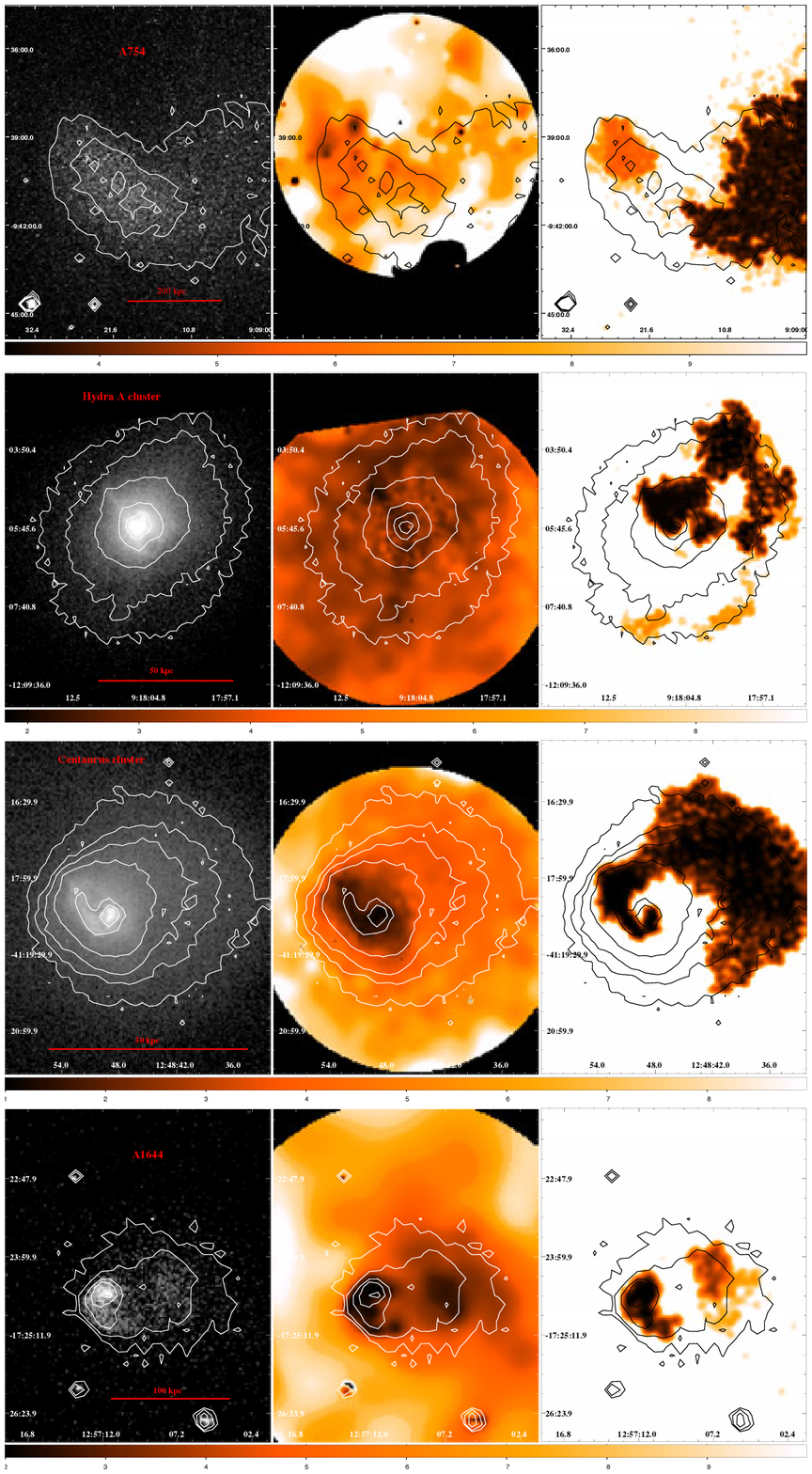}
\caption{{\it Cont.} Left panels: X-ray image; middle panel: temperature maps; and right panels: substructure maps for
A754, Hydra A cluster, Centaurus cluster, and A1644.The color-bar indicates the temperature in keV.}
\end{figure*}

\begin{figure*}[ht!]
\centering
\ContinuedFloat
\includegraphics[width=\tam\textwidth]{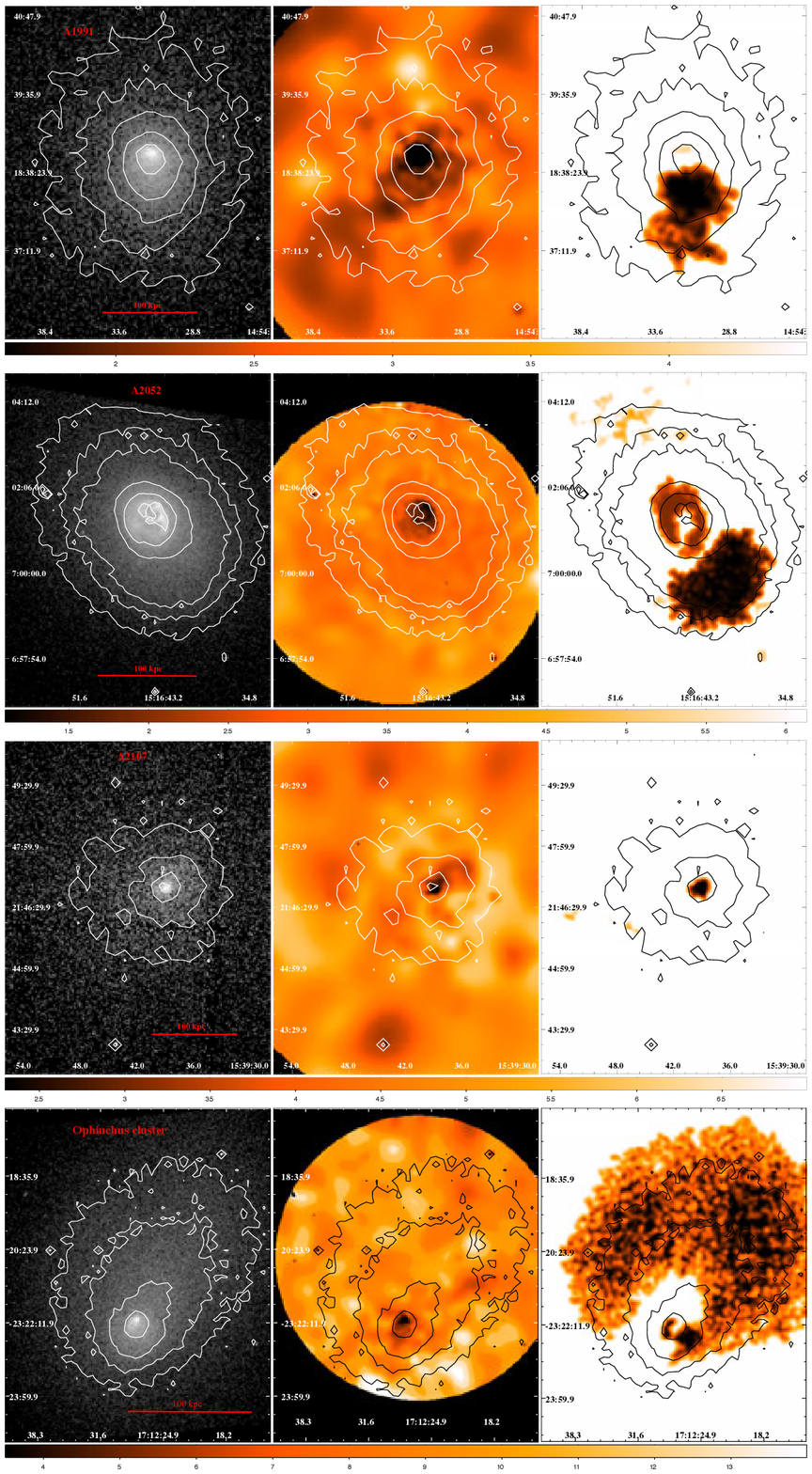}
\caption{{\it Cont.} Left panels: X-ray image; middle panel: temperature maps; and right panels: substructure maps for
A1991, A2052, A2107, and Ophiuchus cluster. The color-bar indicates the temperature in keV.}
\end{figure*}

\begin{figure*}[ht!]
\centering
\ContinuedFloat
\includegraphics[width=\tam\textwidth]{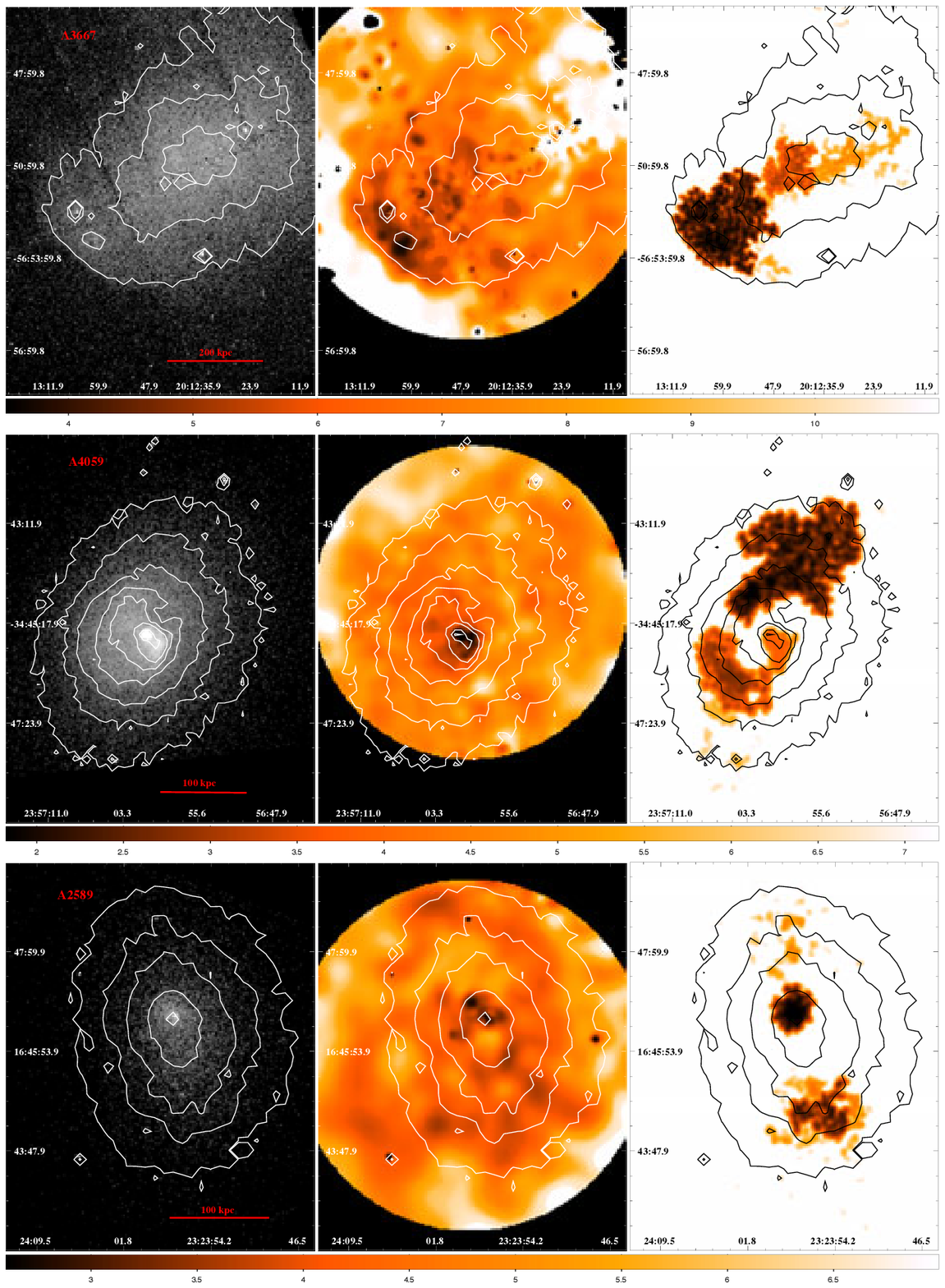}
\caption{{\it Cont.} Left panels: X-ray image; middle panel: temperature maps; and right panels: substructure maps for
A3667, A4059, and A2589.The color-bar indicates the temperature in keV.}
\end{figure*}

\begin{table*}[ht!]
\tiny
\centering
\caption{Cluster properties}
\begin{tabular}{c|cc|ccc|cc|cc}
\hline\hline
 & \multicolumn{2}{c}{ K. Arnaud$^\dag$} & \multicolumn{3}{c}{\citet{Chen07}$^\dag$} & \multicolumn{2}{c}{\citet{Ikebe02}$^\dag$} &  \multicolumn{2}{c}{\citet{Edge92}$^\dag$} \\
 % \cline{2-8}
\hline 
  Cluster 	&  kT & $\dot{M}$  &  kT  & $\dot{M}$  & $L_{X}[\rm 0.1-2.4 keV]$ & kT & $L_{X}[\rm 0.1-2.4 keV]$ &  $L_{X}[\rm 2-10 keV]$ & $\dot{M}$  \\
		& (keV)& ($M_{\odot}~\rm yr^{-1}$) & (keV) & ($M_{\odot}~\rm yr^{-1}$) & $(\rm h_{70}^{-2}~10^{44} \rm erg/s)$ & (keV) & $(\rm h_{70}^{-2}~10^{44} \rm erg/s)$ &  $(\rm h_{70}^{-2}~10^{44} \rm erg/s)$ & ($M_{\odot}~\rm yr^{-1}$) \\
\hline
Abell 85 	&  6.8  &  128.5& $6.10_{-0.20}^{+0.20}$ &  $200_{-27}^{+33}$	 & 13.54 $\pm$ 0.15& $6.51_{-0.23}^{+0.16}$ & 1.81 $\pm$ 0.02 & 10.56  & 236 \\
Perseus (A426) 	&  6.4	&  94.9 & $6.79_{-0.12}^{+0.12}$ &  $481_{-32}^{+31}$	 & 	 --	   & $6.42_{-0.06}^{+0.06}$ & 2.93 $\pm$ 0.04 & 15.40  & 183 \\
Abell  496     	& 5.7 	& 100.7 & $4.13_{-0.08}^{+0.08}$ &  $114_{-28}^{+35}$	 & 5.28 $\pm$ 0.07 & $4.59_{-0.10}^{+0.10}$ & 0.69 $\pm$ 0.01 & 3.56 & 112 \\
Abell 3376     	& -- 	& -- 	& $4.30_{-0.60}^{+0.60}$ &  $0_{-0}^{+0}$	 & 3.02 $\pm$ 0.07 & $4.43_{-0.38}^{+0.39}$ & 0.74 $\pm$ 0.02 & --  & -- \\
Abell  754     	& -- 	& --	& $9.00_{-0.50}^{+0.50}$ &  $0_{-0}^{+0}$	 & 5.56 $\pm$ 0.15 & $9.00_{-0.34}^{+0.35}$ & 4.0 $\pm$ 0.10 & 14.70  & 24   \\
Hydra A       	& -- 	& --	& $3.80_{-0.20}^{+0.20}$ &  $293_{-84}^{+50}$	 & 8.18 $\pm$ 0.06 &  -- 		    & -- 	      & 3.89  & 315  \\
Centaurus 	& 2.3 	& 31.3 	& $3.68_{-0.06}^{+0.06}$ &  $24_{-5}^{+6}$	 & 1.67 $\pm$ 0.06 & 	--		    &	-- 	      & 0.81 & 18\\
Abell 1644      & 5.1 	& 19.3 	& $4.70_{-0.70}^{+0.90}$ &  $0_{-0}^{+0}$	 & 5.49 $\pm$ 0.48 & $4.70_{-0.70}^{+0.90}$ & 0.73 $\pm$ 0.06 & 3.33  & 19 \\
Abell 1991     	& 4.6 	& 71.0 	& --			 &  --  		 & --  		   & & -- & --    & --  \\
Abell 2052     	& 3.7 	& 30.8 	& $3.03_{-0.04}^{+0.04}$ &  $108_{-49}^{+188}$   & 3.32 $\pm$ 0.06 & $3.12_{-0.09}^{+0.10}$ & 0.43 $\pm$ 0.01 & 1.97  & 90  \\
Abell 2107     	& 6.5 	& 0.0 	& -- 			 &  --  		 & --  		   & & -- & --    & --  \\
Ophiuchus	& -- 	& -- 	& $10.26_{-0.32}^{+0.32}$&  $0_{-0}^{+0}$	 & --  		   & $10.25_{-0.36}^{+0.30}$ & 2.2 $\pm$ 0.07 & 21.28  & 75\\
Abell 3667     	& -- 	& -- 	& $7_{-0.60}^{+0.60}$	 &  $0_{-0}^{+0}$	 & 13.27 $\pm$ 0.15 & $6.28_{-0.26}^{+0.27}$ & 1.78 $\pm$ 0.02 & 11.59  & 0 \\
Abell 4059     	& -- 	& -- 	& $4.10_{-0.30}^{+0.30}$ &  $69_{-15}^{+20}$	 & 3.92 $\pm$ 0.08 & $3.94_{-0.15}^{+0.15}$ & 0.52 $\pm$ 0.01 & 2.65  & 124\\
Abell 2589     	& -- 	& --	& $3.70_{-1.10}^{+2.20}$ &  $19_{-19}^{+53}$	 & 2.62 $\pm$ 0.06 & $3.38_{-0.13}^{+0.13}$ & 3.44 $\pm$ 0.08 & --    & -- \\

\hline
\hline	  
\end{tabular}
\\
\begin{minipage}{10.0truecm}
\tiny{$^\dag$ All values were converted to a common cosmology with a Hubble constant given as $\rm H_{0}=70$~km s$^{-1}$Mpc$^{-1}$.}\\
\end{minipage}
\label{prop}
\end{table*}

Although not clearly evident in all temperature maps, a spiral-like structure in the inner part of 7 clusters 
(A85, A426, A496, Hydra A cluster, Centaurus, Ophiuchus, and A4059) is detected in the
substrucuture maps. These clusters are those of the highest cooling-flow rate and most are
known to have a strong radio galaxy at the centre.

Since the variations in temperature and density inside shocks are positive correlated, 
that these patterns are also
regions of low temperature and high density argues against strong shocks.
These substructures instead, resemble the edges of cold fronts found in several clusters \citep[e.g.,][]{Mark07,Dupke07}.
Cold fronts can be directly related to cluster merging activities, and by comparing our results with 
numerical simulations \citep{Gomez02,Ascasibar06} one can see that the spiral-like features found here are morphologically similar to those found in their simulations. 
This indicates that their nature is probably dynamical and a possible scenario 
for its formation is discussed in Sect.~\ref{disc}.

Here, we provide additional information about each cluster in the sample, 
by highlighting key aspects of previous studies.

\subsection{Abell 85}
This is a well-known subclump cluster that has a cold-front \citep{Kempner02} and is a merger candidate.
The optical properties of this cluster are indicative of evidence for galaxy star formation in the filament between the main cluster and the subclump \citep{Boue08}, 
which is consistent with the previous hypothesis that the 
X-ray filament in Abell 85 is a gravitationally bound structure consisting of groups falling 
into the main cluster \citep{LN01,Durret03}.

We clearly see a spiral-like structure in the substructure map. However, 
no corresponding structure appears in the temperature map.

\subsection{Abell 426 - Perseus}

\citet{Churazov03} presented XMM-\textit{Newton} data in which a spiral-like structure 
was detected in temperature and surface
brightness maps. These authors concluded that this feature is possibly a contact discontinuity separating the main cluster gas from the gas of the infalling subcluster. A chain of galaxies is also associated with this 
region, probably tracing the filament along which the merger started. 

Since it has a strong central radio galaxy, NGC 1275, \citet{Bohringer93} argued that the 
thermal plasma was displaced by the inner parts of the radio lobe.
\citet{Fabian02} examined in detail the interaction between the radio source 3C84 and
the surrounding medium, arguing that the inner lobes are currently expanding subsonically.

From our results, one can see that this cluster is one with the most prominent spiral-arm patterns
in both temperature and substructure maps. The substructure found here is almost identical to that
detected by \citet{Churazov03} (see their Fig. 7).

\subsection{Abell 496}
This cluster hosts a prominent cold-front \citep{Dupke03}, which is a sign of a recent minor merger, 
and although its resport has not received much attention, a very prominent spiral-like pattern
was detected by \citet{Lagana08} in XMM-\textit{Newton} data.

In Fig.~\ref{HR}, a spiral-like feature is evident in both the substructure and temperature map.
The pattern that appears in the temperature map is very similar in form to that of A85.

\subsection{Abell 3376}
This is a nearby ongoing merger cluster that was bserved by both \textit{Suzaku} \citep{Kawano08} and 
XMM-\textit{Newton}
\citep{Bagchi06}. A non-thermal radio-emitting structure at the outskirts of this cluster was identified 
by the latter authors. They suggested that this structure probably traces the elusive shocks of cosmological 
large-scale matter flows.

\subsection{Abell 754}
A754 was one of the first clusters for which an X-ray temperature map was derived from ROSAT 
\citep{Henry95} and ASCA \citep{Henriksen96} data, in which the gas velocity map \citep{Evrard96} 
indicated a plume-like pattern. A {\it Chandra} data analysis inferr a more complex merger geometry, 
possibly involving more than two subclusters \citep{Mark03}. From XMM-\textit{Newton} data, 
\citet{Kassim01} and \citet{Bacchi03} confirmed that this cluster is a major-merger cluster and also detected 
a plume-like feature at its centre.
There is a diffuse radio source associated with this cluster \citep{Henry04}. 

In this work, a large substructure is detected but does not exhibit any particular shape.

\subsection{A780 - Hydra A cluster}
\citet{Fitchett88} examined the dynamics of the Hydra cluster find that the velocity 
distribution of galaxies close the cluster centre is very flat. 
Although they did not propose an infalling group scenario, 
they did consider the importance of substructures in this cluster, 

{\it Chandra} observations of the Hydra cluster detect a feature in the X-ray surface brightness surrounding 
radio lobes of the active galactic nucleus (AGN) at 
the cluster centre \citep{Nulsen05}. 
These authors proposed that the shock front is driven by the expanding radio lobes in addition to the ``new 
structure on smaller scales'' is associated with the radio source.
\citet{Wise07} confirmed that the complex system found in this cluster was created by a continuous 
outflow or a series of bursts in the nucleus of the central galaxy.
In this particular case, a merger has not been mentioned in the literature as a possible 
mean of forming this substructure.

In this work, we were able to detect a truncated spiral-like feature in the 
substructure map that has a 
corresponding structure of relatively low temperature.

\subsection{Centaurus}
To explain the particular morphology of this cluster,
\citet{Churazov99} analysed ASCA imaging and spectral results, 
and proposed that there is a subcluster
centred on NGC 4709 that is merging with the main cluster centred on NGC 4696.

\citet{Sanders06} performed a detailed analysis of the core of the Centaurus
cluster using a {\it Chandra} deep X-ray observation and XMM-{\textit Newton} data.
To explain the high metallicity present in the core of this cluster, 
the authors proposed a model in which the inner core of the Centaurus cluster 
has not experienced a major disruption within the past 8 Gyr, or even longer. 
In their temperature map, the plume-like feature is evident, 
this finding then attracted
little attention.
They also reported the discovery of 
ripple-like X-ray surface brightness oscillations 
in the core of this cluster of galaxies, which they asserted were indicative of sound waves 
generated by the repeated inflation of central radio bubbles.

From Fig.~\ref{HR}, one can see that this cluster has one of the most
prominent spiral-like features detected in a substructure map.
The very inner region of this enormous structure is also associated with a region
of relatively low temperature in the temperature map.

\subsection{Abell 1644}
This is a complex merging system that was observed with XMM-\textit{Newton}, and found to consist of
a main cluster and a subcluster \citep{Reiprich04}. 
These authors also detected a trail of cool, metal-rich gas closer to the subcluster. 
The combination of X-ray, optical, and radio data results imply that the subcluster 
has passed by the main cluster off-axis. Although temperature maps were presented in their work, 
the plume-like feature was not resolved by XMM-\textit{Newton}.

In our analysis, a discontinuous spiral-like feature is present in the substructure map.
This pattern is not as clearly evident as in Centaurus, Perseus, and A85 but it does resemble these structures. 
Unfortunately, only the inner part of the substructure has a corresponding feature in the temperature map.

\subsection{Abell 1991}
\citet{Sharma04} analysed XMM-{\textit Newton} data and detected an asymmetric surface brightness 
distribution with respect to the central galaxy. 
These authors detected bright knots of soft X-ray emission 
embedded in a cometary structure located north of the optical centre of the cD galaxy. 
According to these authors, the knots have no obvious association with the radio source.

\subsection{Abell 2052}
\citet{Zhao93} reported a VLA detection of a radio source at the core of this cluster. 
\citet{Blanton03,Blanton09} carried out a {\it Chandra} X-ray data analysis of the large-scale properties of the
cluster as well as the central region (which contains a radio source) presenting temperature
and abundance profiles.

Two clear substructures are present in Fig.\ref{HR} but they do not resemble to the
other spiral-like feature presented in this work.

\subsection{Abell 2107}
\citet{Fujita06} presented an analysis of \textit{Chandra} observation of this cluster. 
Since the cD galaxy has a large peculiar velocity and the ICM in the central region 
has an irregular structure, these authors concluded that this cluster is undergoing a merger.

From our results, A2107 contains the smallest substructure and has almost symmetric X-ray contours
providing no evidence of a recent merger.

\subsection{Ophiuchus}
Ophiuchus is one of the hottest X-ray cluster and is located 12 deg from the Galactic centre \citep{Fujita08}.
Using ASCA data, \citet{Matsu96} acquir images in the energy range of 0.7-10 keV that indicated that the 
peak of the X-ray surface brightness is coincident with the cD galaxy. \citet{Watanabe01} presented 
temperature and metallicity profiles obtained with ASCA data, which contained no evidence of a plume-like feature.
In their analysis, the ROSAT archival data inferred an ongoing merger, but a \textit{Chandra} image 
indicates that this cluster has a cool-core that remains undisturbed. Thus,
if there is an ongoing merger, it must be a minor merger \citep[e.g.,][]{Watanabe01}.

From the results presented in Fig.\ref{HR}, Ophiuchus is the cluster that has 
the most prominent spiral-like pattern in its substructure map.
However, no significant feature was detected in its temperature map.

\subsection{Abell 3667}
This cluster has a double radio halo outside its central region \citep{Rott97,Mark99}.
The galaxy distribution is bimodal \citep{Proust88}, the main component being located
around the cD galaxy. A multiwavelength study confirmed that it is probably a merger cluster
based on \textit{Chandra} and optical data \citep{Owers09b}.
Its X-ray image shows a clear cold front.

\subsection{Abell 4059}
\citet{Heinz02} affirmed that there are clear signs of interaction between 
the radio galaxy and the ICM. 
However, these authors did not find any one-to-one spatial correspondence
between the radio lobes and the X-ray cavities \citep{Heinz02}.
\citet{Huang98} argued that it is likely that the radio plasma from the cD galaxy 
may have displaced the X-ray gas and created the X-ray cavities.

On the other hand, \citet{Choi04} analysed various scenarios concluding that 
the most probable is that of a compression front associated with ICM bulk motion.

\citet{Reynolds08} analysed a \textit{Chandra} deep observation, 
detect a subtle discontinuity in the gradient of the surface brightness
that is approximately semi-circular in form and centred on the radio galaxy.
The previous \textit{Chandra} observations reported by \citet{Heinz02} 
and \citet{Choi04} were of insufficient signal-to-noise ratio to detect this feature. 
Although it cannot be rigorously proven,
\citet{Reynolds08} interpreted this feature 
as a weak shock caused by the radio galaxy activity. Similar shocks are
seen in all hydrodynamic simulations of jet/ICM interactions
\citep[e.g.]{Heinz02,Vernaleo06}.

In this work, we have shown that this cluster also has a spiral-like substructure.
The inner part of this pattern is related to
a region of lower temperature, as it can be seen in the temperature map of Fig.\ref{HR}.

\subsection{Abell 2589}
The hot gas in the core region of A2589 is undisturbed by interactions with 
a central radio source \citep{Buote04}. 
Except for a ~16 kpc shift in its X-ray centre, A2589 has a 
remarkably symmetrical X-ray image \citep{Zappa06}.
 
\section{Discussion}
\label{disc}

The literature contains detailed studies of large numbers of clusters with high
cooling rates and a powerful radio galaxy, e.g., Hydra A \citep{McNamara00,David01},
Perseus \citep{Fabian00, Churazov03}, and the Centaurus cluster \citep{Taylor94}.
There is a correlation between the radio emission by the central galaxies and the presence of
a cooling-flow that has been known since \citet{Burns81} and \citet{JF84}.
The cooling gas is driven towards a massive object in the central galaxy that might produce
the radio emission. However, what is quite surprising is that clusters exhibiting distinctive 
spiral-like structures are those with the greatest 
cooling-flows (see Table~\ref{prop}) and containing a powerful radio galaxy. 
The only exception is A2052 that contains an AGN \citep{Blanton03,Blanton09} 
and does not exhibit a spiral-like substructure. 

The presence of a strong cooling-flow and a powerful central radio galaxy indicates that
its radiation has not yet had time to heat the surrounding gas of the inner parts of the clusters.
This can be understood if we consider a cycle in which the cool gas feeds the central
AGN that begins to heat the surrounding gas. While the central gas is not sufficiently 
heated and the cooling flow 
is not suppressed, there will be a stage where both the AGN and the cool gas will coexist.

The radio morphology of some clusters that exhibit spiral-like patterns appears quite different
from most radio galaxies because they fill the X-ray cavities (regions of low density). 
\citet{Taylor02} showed that the eastern lobe of the central radio 
galaxy of Centaurus cluster appears to fill the region bounded by the spiral-like structure,
while the western lobes turns around it. Similar confinement has been found for other radio
galaxies embended in dense ICM. For example, radio emission filling X-ray cavities was found 
at the core of Perseus cluster \citep{Bohringer93,Fabian00,Fabian02}, 
A2029 \citep{Taylor94}, and Hydra A cluster (A780) \citep{McNamara00}.

The ICM of merging clusters exhibit a variety of signatures of mergers such as
cold-fronts and bow shocks. 
Although the most prominent cold fronts are produced in major mergers, 
cold fronts produced by minor mergers are expected to be far more frequent.
Minor mergers generate subsonic turbulent gas motions within the cluster without destroying the cool-core.
Different morphological features can be created by of minor mergers, such as
comet-like tails, bridges, plumes, and edges \citep{Poole06}. 
Thus, one possibility for the origin of the spiral structure found in 7 clusters of our sample
is that they are the signature of a small galaxy cluster or group minor merger with
a more massive cluster. 

In a high-resolution numerical simulation of idealized cluster mergers, 
\citet{Ascasibar06} concluded that if an infalling subcluster has a non-zero impact parameter, 
the cool gas of the main cluster acquires angular momentum, producing in a spiral-like feature of cold front 
due to the gas sloshing. In their simulation, after 1.9 Gyr 
(see their Fig.~7) one can discern a spiral-like pattern in both 
the gas density and temperature distribution,
that are morphologicaly similar to the structure found here.
Since they used a mass ratio of 1:5, the cooling core is preserved. 

The numerical simulations of \citet{Gomez02} also reproduce the results found here.
In their merger 7, the subcluster has a small gas fraction and they assume a mass ratio of 1:16 
so that the cooling-flow is not disrupted during the merger. An alternative explanation of the spiral-like  
structure is a contact discontinuity separating the main cluster gas from the gas of the infalling subcluster.
However, it is also possible that the gas is stripped way from the infalling subcluster far from the core, 
never reaching the inner region of the main cluster \citep{Churazov03}.
In this way, the observed strucutre consists only of the disturbed gas of the main cluster.
A metallicity analysis would be able to distinguish between the above scenarios. 
If the cold gas in the spiral-like
pattern is caused by the main cluster gas sloshing, the metallicity in this region should be equal 
to the metallicity in the inner part of the main cluster. On the other hand, if it is a gas discontinuity, 
the metallicity in the spiral arm would be different from the inner part of the main cluster. 
The ICM enrichment analysis for this sample will be presented in a forthcoming work.

In minor merger scenario, it is more likely that the cooling flow continues to feed the central galaxy. 
The radio emission then, permeates the regions of low density, that is, 
fills the X-ray cavities because it is confined by the spiral-like feature. 
The ICM acts to distort the radio emission producing the
radio morphologies, as can be seen in some clusters.

We note that the detection of spiral-like features is not limited to the clusters 
in our sample but has also been achieved in other clusters, such as A1975 \citep{Fabian01}, 
A262 \citep{Blanton04},
A2029 \citep{Clarke04}, A5098 \citep{Randall09}, A1795 \citep{Liuzzo09}, and A1201 \citep{Owers09}.
This kind of pattern has also been observed in galaxy groups \citep{Gastaldello07,Randall09},
indicating that they are not restricted to clusters.

Since we find that a high number of clusters contain these structures, 
it would be interesting to bolster these findings by using
numerical simulations. In particular, it would be instructive to study the frequency of off-axis merger 
halos, as a function of redshift, that produce a cool gas spiral-like pattern when adopting
mass ratios of 1:5 \citep{Ascasibar06} and 1:16 \citep{Gomez02}.  
Unfortunately, these studies are beyond the scope of our present investigation.

From the theoretical point of view, future numerical hydrodynamical simulations could
detect the signatures of cluster mergers that could then be compared with observations.
In this sense, statistical studies of cluster morphology and substructures would provide 
important constraints on models of structure formation and the evolution of the observed cluster properties.
Despite the large amount of available data, the dynamical evolution of ICM substructures
remains poorly understood. Here we have studied in detail 15 nearby galaxy clusters highlighting 
the incidences of a particular
ICM substructure morphology.

\section{Synthetic image}
\label{SyntImg}
\begin{figure}[ht!]
\centering
\includegraphics[width=0.5\textwidth]{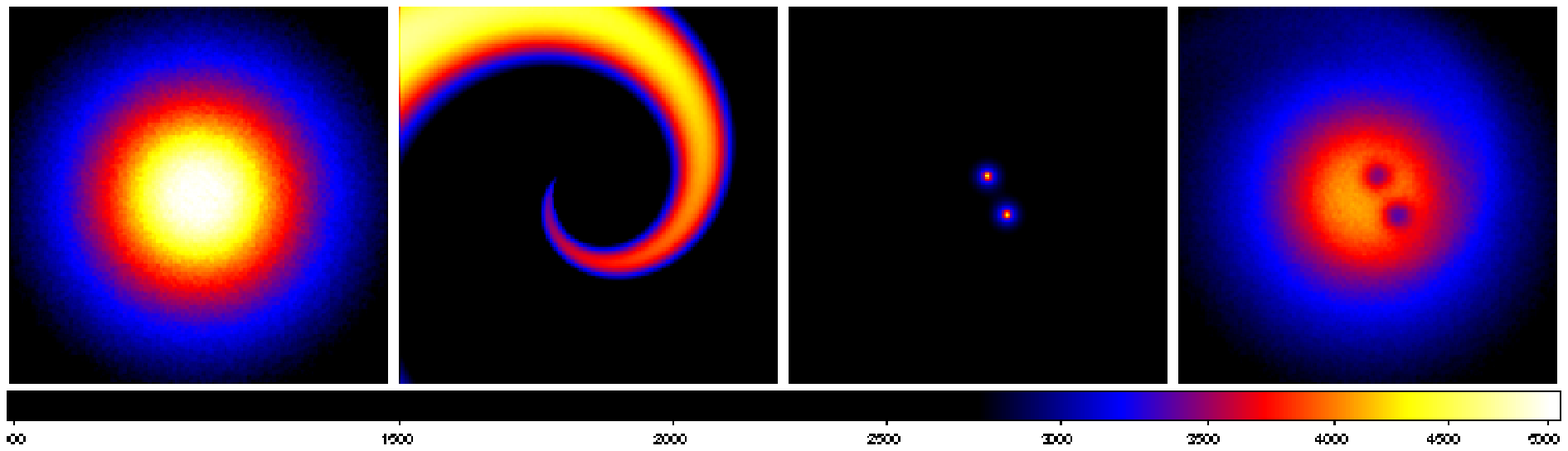}
\includegraphics[width=0.5\textwidth]{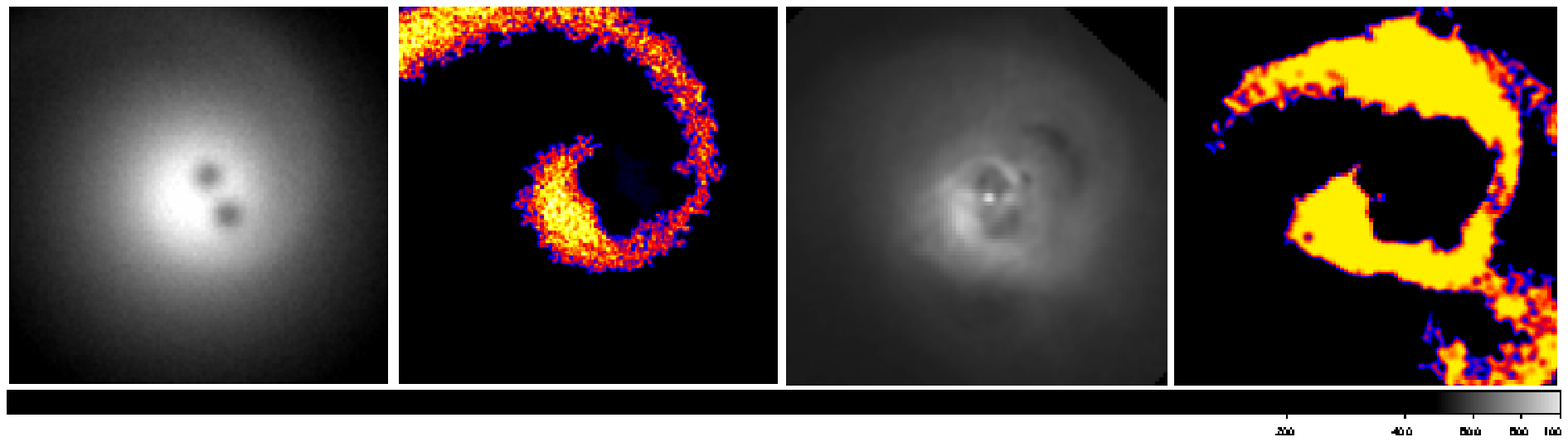}
\caption{Synthetic image constructed to roughly reproduce the Perseus cluster X-ray surface brightness.
\textsf{Top panels from left to right}: 2D $\beta$-model, exponential spiral pattern, two bubbles, 
and the sum of these
three components. \textsf{Bottom panels from left to right}: X-ray synthetic surface brightness, 
substructure map from this
previous image, Perseus X-ray surface brightness, and its substructure map.}
\label{syn_img}
\end{figure}

We now present a synthetic image, created using the
\textit{cfitsio} routine (a library of routines for reading and writing data
files in the FITS data format).
This synthetic image was constructed by adding multiple components to
roughly mimic the Perseus cluster X-ray surface brightness.
The synthetic image was the result of adding a 2D $\beta$-model, an
exponential spiral pattern, and two bubbles in the inner region of the
cluster.
The components and the result of their addition is seen in Fig. \ref{syn_img}.

The goal of generating this synthetic image is to follow the same procedures
applied  to true clusters, on determining substructure, and comparing
with the substructure of the Perseus cluster. Although clearly present, one can
see that the spiral pattern is not distinctive in the synthetic surface
brightness image (see top panel of Fig.~\ref{syn_img}), which shows the
importance of this method in finding structures that are hardly discernible by eye.

The synthetic-image substructure map shows the result obtained in  a
spiral-like pattern structure very similar to the one found in the Perseus
clusters, reinforcing the detectability of these spiral-like patterns.
Furthermore, the result shows that the Perseus cluster X-ray surface
brightness and substructures can be described well by the addition
of three components, to produce in particular an exponential spiral-like pattern,
similar to that found in the results of numerical simulations performed 
by \citet{Ascasibar06}.

\section{Conclusions}
\label{conc}

In the most widely accepted scenario of structure formation, clusters of galaxies form hierarchically 
by means of mergers. Besides the modification of galaxy distribution and
triggering of star formation, these merger events produce variations in the physical properties, 
such as density and temperature variations that can be observed in X-rays.
In this sense, the substructure maps have provided a powerful means 
of identifying hidden structures that are the last record of the merger history.

We have presented 7 (A85, A426, A496, Hydra A cluster, Centaurus, Ophiuchus and A4059) 
out of 15 clusters that exhibit for spiral-like substructure correlated spatially with the 
boundaries of cold-fronts and cavities.
The cavities inside these patterns are, in most of cases, associated with radio emission.
The aim of this study has been to determine the frequency of this morphological 
pattern at the centres of nearby clusters.
Our analysis has shown how observational studies can improve the constraints on numerical simulations of
structure formation and improve our understanding of ICM physical properties.

Comparing our results to numerical simulations \citep{Ascasibar06},
we have discussed the suitability of a model in which this spiral-like pattern is produced by
an off-axis minor-merger that displaces the central cool gas. 
Since the minor merger does not disrupt the cool core, 
the central cool gas is still driven towards the central object that may produce 
the radio emission. 
This radio emission is bounded by the ICM inhomogeneities that will 
probably create an overall radio appearance that is quite 
different from most radio galaxies. 
Thus, we suggest that these distorted radio-emission
morphologies may be caused by the confinement 
of dense X-ray gas in the spiral pattern.
However, our interpretation of the spiral-like pattern and its relation 
to X-ray cavities and radio emission should be tested by additional more sophisticated 
simulations that ascertain possible merger signatures.

\begin{acknowledgements}
The authors thank Florence Durret that triggered this work.
The authors thank the anonymous referee for usefull suggestions.
The authors also acknowledge financial support from the Brazilian 
agencies FAPESP (grants: 2006/56213-9, 2008/04318-7, and 2008/05970-0) 
and CNPq (grant: 472012/07-0).
\end{acknowledgements}

\bibliographystyle{aa}
%
%  These Macros are taken from the AAS TeX macro package version 4.0.
%  Include this file in your LaTeX source only if you are not using
%  the AAS TeX macro package and need to resolve the macro definitions
%  in the BibTeX entries returned by the ADS abstract service.
%
%  If you plan not to use this file to resolve the journal macros
%  rather than the whole AAS TeX macro package, you should save the
%  file as ``aas_macros.sty'' and then include it in your paper by
%  using a construct such as:
%	\documentstyle[11pt,aas_macros]{article}
%
%  For more information on the AASTeX macro package, please see the URL
%	http://www.aas.org/publications/aastex.html
%  For more information about ADS abstract server, please see the URL
%	http://adswww.harvard.edu/ads_abstracts.html
%

% Abbreviations for journals.  The object here is to provide authors
% with convenient shorthands for the most "popular" (often-cited)
% journals; the author can use these markup tags without being concerned
% about the exact form of the journal abbreviation, or its formatting.
% It is up to the keeper of the macros to make sure the macros expand
% to the proper text.  If macro package writers agree to all use the
% same TeX command name, authors only have to remember one thing, and
% the style file will take care of editorial preferences.  This also
% applies when a single journal decides to revamp its abbreviating
% scheme, as happened with the ApJ (Abt 1991).

\def\ref@jnl{}

\def\aj{\ref@jnl{AJ}}                   % Astronomical Journal
\def\araa{\ref@jnl{ARA\&A}}             % Annual Review of Astron and Astrophys
\def\apj{\ref@jnl{ApJ}}                 % Astrophysical Journal
\def\apjl{\ref@jnl{ApJ}}                % Astrophysical Journal, Letters
\def\apjs{\ref@jnl{ApJS}}               % Astrophysical Journal, Supplement
\def\ao{\ref@jnl{Appl.~Opt.}}           % Applied Optics
\def\apss{\ref@jnl{Ap\&SS}}             % Astrophysics and Space Science
\def\aap{\ref@jnl{A\&A}}                % Astronomy and Astrophysics
\def\aapr{\ref@jnl{A\&A~Rev.}}          % Astronomy and Astrophysics Reviews
\def\aaps{\ref@jnl{A\&AS}}              % Astronomy and Astrophysics, Supplement
\def\azh{\ref@jnl{AZh}}                 % Astronomicheskii Zhurnal
\def\baas{\ref@jnl{BAAS}}               % Bulletin of the AAS
\def\jrasc{\ref@jnl{JRASC}}             % Journal of the RAS of Canada
\def\memras{\ref@jnl{MmRAS}}            % Memoirs of the RAS
\def\mnras{\ref@jnl{MNRAS}}             % Monthly Notices of the RAS
\def\pra{\ref@jnl{Phys.~Rev.~A}}        % Physical Review A: General Physics
\def\prb{\ref@jnl{Phys.~Rev.~B}}        % Physical Review B: Solid State
\def\prc{\ref@jnl{Phys.~Rev.~C}}        % Physical Review C
\def\prd{\ref@jnl{Phys.~Rev.~D}}        % Physical Review D
\def\pre{\ref@jnl{Phys.~Rev.~E}}        % Physical Review E
\def\prl{\ref@jnl{Phys.~Rev.~Lett.}}    % Physical Review Letters
\def\pasp{\ref@jnl{PASP}}               % Publications of the ASP
\def\pasj{\ref@jnl{PASJ}}               % Publications of the ASJ
\def\qjras{\ref@jnl{QJRAS}}             % Quarterly Journal of the RAS
\def\skytel{\ref@jnl{S\&T}}             % Sky and Telescope
\def\solphys{\ref@jnl{Sol.~Phys.}}      % Solar Physics
\def\sovast{\ref@jnl{Soviet~Ast.}}      % Soviet Astronomy
\def\ssr{\ref@jnl{Space~Sci.~Rev.}}     % Space Science Reviews
\def\zap{\ref@jnl{ZAp}}                 % Zeitschrift fuer Astrophysik
\def\nat{\ref@jnl{Nature}}              % Nature
\def\iaucirc{\ref@jnl{IAU~Circ.}}       % IAU Cirulars
\def\aplett{\ref@jnl{Astrophys.~Lett.}} % Astrophysics Letters
\def\apspr{\ref@jnl{Astrophys.~Space~Phys.~Res.}}
                % Astrophysics Space Physics Research
\def\bain{\ref@jnl{Bull.~Astron.~Inst.~Netherlands}} 
                % Bulletin Astronomical Institute of the Netherlands
\def\fcp{\ref@jnl{Fund.~Cosmic~Phys.}}  % Fundamental Cosmic Physics
\def\gca{\ref@jnl{Geochim.~Cosmochim.~Acta}}   % Geochimica Cosmochimica Acta
\def\grl{\ref@jnl{Geophys.~Res.~Lett.}} % Geophysics Research Letters
\def\jcp{\ref@jnl{J.~Chem.~Phys.}}      % Journal of Chemical Physics
\def\jgr{\ref@jnl{J.~Geophys.~Res.}}    % Journal of Geophysics Research
\def\jqsrt{\ref@jnl{J.~Quant.~Spec.~Radiat.~Transf.}}
                % Journal of Quantitiative Spectroscopy and Radiative Transfer
\def\memsai{\ref@jnl{Mem.~Soc.~Astron.~Italiana}}
                % Mem. Societa Astronomica Italiana
\def\nphysa{\ref@jnl{Nucl.~Phys.~A}}   % Nuclear Physics A
\def\physrep{\ref@jnl{Phys.~Rep.}}   % Physics Reports
\def\physscr{\ref@jnl{Phys.~Scr}}   % Physica Scripta
\def\planss{\ref@jnl{Planet.~Space~Sci.}}   % Planetary Space Science
\def\procspie{\ref@jnl{Proc.~SPIE}}   % Proceedings of the SPIE

\let\astap=\aap
\let\apjlett=\apjl
\let\apjsupp=\apjs
\let\applopt=\ao
 
\bibliography{SA_corrected}

\end{document}